\documentclass[10pt,letterpaper]{article}
\usepackage{opex3}
\usepackage{subfigure}
\usepackage{cite}

\begin{document}

\title{Dark-field digital holographic microscopy for 3D-tracking of gold nanoparticles}

\author{F.~Verpillat,\textsuperscript{1*} F.~Joud,\textsuperscript{1} P.~Desbiolles,\textsuperscript{1} and M.~Gross~\textsuperscript{2}}

\address{\textsuperscript{1} Laboratoire Kastler Brossel, ENS, UPMC-Paris6, CNRS UMR 8552, Paris, 75005, France}
\address{\textsuperscript{2} Laboratoire Charles Coulomb, CNRS UMR 5221, Universit\'e Montpellier II, Montpellier, 34095, France}

\email{frederic.verpillat@lkb.ens.fr}

\begin{abstract}
We present a new technique that combines off-axis Digital Holography and Dark Field Microscopy to track $100\, \rm nm$ gold particles diffusing in water. We show that a single hologram is sufficient to localize several particles in a thick sample with a localization accuracy independent of the particle position. From our measurements we reconstruct the trajectories of the particles and derive their 3D diffusion coefficient. Our results pave the way for quantitative studies of the motion of single nanoparticle in complex media.
\end{abstract}



\section{Introduction}
The study of cellular processes at the single-molecule level is a flourishing field of research in Biology. Individual molecules labeled with sub-micron markers can now be tracked in a cellular environment, and quantitative information about their dynamics can be obtained by reconstructing their trajectory. One of the most used techniques for this purpose is single-molecule fluorescence microscopy (SMFM), which relies on a labeling with nanometer-sized fluorescent markers such as organic dyes or quantum dots. But standard SMFM provides no information on the axial position of the marker, limiting this technique to 2D tracking. Recent improvements of SMFM such as astigmatism optic\cite{huang2008three}, $4$Pi microscopy \cite{bewersdorf2006h2ax}, double-helix PSF \cite{pavani2009three}, or multi-planes detection \cite{juette2008three,ram2008high} have made possible 3D tracking. Since the depth of field of these techniques is limited to a few microns, 3D tracking of molecules that explore larger distances in the thickness of a sample requires to continuously adjust the position of the focal plane of the microscope objective, which strongly limits the time resolution. Digital Holographic Microscopy (DHM) \cite{schnars1994drh,leith1965} circumvents this drawback. In DHM, a CCD camera records the interference pattern between the light scattered by the sample and a reference wave, and a single shot is sufficient to determine the 3D positions of scatterers embedded in a non-diffusing environment, over a depth of typically a hundred of microns.

As the scattering cross section of a particle scales as the sixth power of its radius \cite{vandeHulst}, how easily and accurately a particle can be detected strongly depends on its size. Several publications demonstrate the tracking of micron-sized colloids by using in-line holography \cite{cheong2009flow,cheong2009holographic,cheong2010strategies,xu2003tracking,sheng2006digital,lee2007characterizing,speidel2009interferometric}, with a localization accuracy in the nanometer range through the use of high Numerical Aperture (NA) microscope objectives. For example, with a $100\times$ NA 1.4 oil immersion objective, Cheong et al. \cite{cheong2009holographic} reported lateral and axial  localization accuracies of 4  and 20 nm respectively. This result was obtained with polystyrene spheres of diameter $d=1.5~\mu$m, whose scattering cross section is quite large. The tracking of
$d \leq 100$ nm particles, whose scattering cross section is extremely low, is much more difficult, and, as far as we know, has not been demonstrated using in-line holography. Yet the detection of such small particles is possible using DHM in a off-axis geometry \cite{leithjosa,cuche2000spatial} with a noise level as low as possible \cite{gross2007digital} and using good light scatterers, such as gold nanobeads \cite{jain2006calculated}. By this way, $d=50$ to $200$ nm gold particles  embedded in an agarose gel have been detected and localized \cite{atlan2008hhm}. Since gold particles are not toxic for cells, they can be used as markers in biology \cite{cognet2003single}, and d = 40 nm gold nanobeads fixed on the membrane receptors of a living cell have been localized \cite{warnasooriya2010imaging}. More recently, 3D tracking of BaTiO3 particles with second harmonic generation DHM was also demonstrated \cite{shaffer2010real}.

Here, the main advantage of DHM, with respect to in-line holography, is the possibility to independently adjust the intensity of the illumination and reference beams in order to get the best
detection sensitivity, by adjusting the reference beam intensity \cite{verpillat2010digital}, and the largest signal, by adjusting the illumination beam intensity. Combining DHM with dark field illumination allows then to detect nanometer-sized  particles, as the sample can be illuminated with an intensity as large as possible, while avoiding a saturation of the camera \cite{atlan2008hhm}. For example, Atlan et al. and Warnasooriya et al. uses Total Internal Reflection (TIR) to detect and localize $d=50$ nm and $d=40$ nm particles \cite{atlan2008hhm,warnasooriya2010imaging}. But the TIR configuration used in these experiments yield a standing wave which does not allow to track moving particles : when a moving particle crosses a node, the illumination (and thus the signal) goes down to zero, and the particle is lost. Dubois et al. uses another dark field illumination configuration that focuses the illumination on a mask \cite{dubois2008dark}. Since the illumination is parallel to the optical axis, no standing wave can appear, but since the illumination passes through the microscope objective, one expects parasitic reflections of the illumination beam.

In this paper, we present a Digital Holographic Microscopy technique which makes possible to track $d=100$ nm gold particles 3D diffusing in water. The illumination is parallel to the optical axis to prevent the formation of standing waves, and the holographic signal is collected by a NA=0.5 dark field reflecting microscope objective. With such objective, the illumination beam is masked before the
microscope objective and parasitic reflections are avoided. This yields high dynamic dark field illumination, which makes possible to detect, localize and track $d=100$ nm particles.
First we describe the setup, which combines dark-field microscopy and off-axis holography. Then we present the algorithm of reconstruction, our procedure to localize the beads, and describe how we can reach a real-time localization by performing calculations on a graphic card. Finally we show that our setup allows us to track gold nanoparticles in motion with a lateral ($x,y$) resolution of $\sim 3$ nm and an axial ($z$) resolution of $\sim 70\ \rm nm$. Since NA=0.5, the resolution (especially in $z$) is lower than with NA=1.4 in-line holography \cite{cheong2009holographic}. We also show that the depth of field of our holographic microscope is made two orders of magnitude larger than in optical microscopy.

\section{Digital holography setup}

\begin{figure}[htbp]
\centering
\includegraphics[width = 10.0 cm]{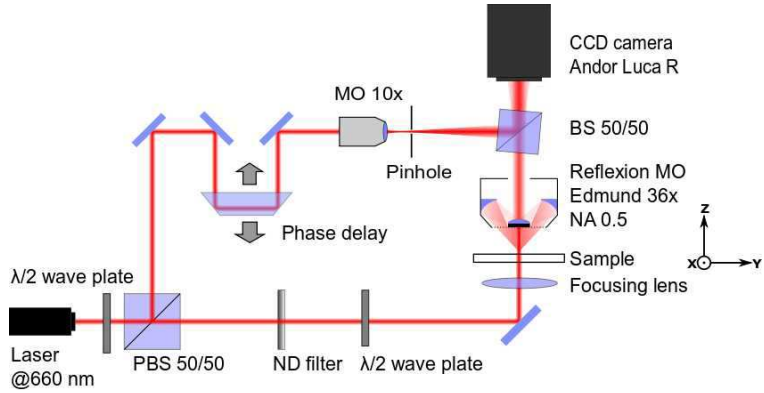}
\caption{Experimental setup. The sample is located in the $X$,$Y$ plane. $Z$ is the optical axis of the microscope objective}
\label{fig:setup}
\end{figure}

Our DHM experimental setup, depicted in Fig. \ref{fig:setup}, is designed to investigate the Brownian motion of $100 \, \rm nm$ diameter gold particles diffusing in a $17 \times 3.8 \times 0.4\, \rm mm$ (length $\times$ width $\times$ height) cell chamber (Ibidi \textcopyright ${\, \rm \mu\!-\!Slide}$) filled with water. The concentration of nanoparticles is adjusted to $5.6\  10^3$ particles/mm$^3$ to have a few particles per field of view.

The light source is a $660\, \rm nm$  Diode Pump Solid State laser (Crystal Laser \textcopyright) with a short coherence length ($\sim 600\, \mu \rm m$) to avoid parasitic interferences raising from reflexions between the optical elements. The laser beam is split into two beams by a Polarizing Beam Splitter (PBS), a half-wave plate before the PBS setting the ratio of energy between the emerging beams. The reference beam passes through a dove prism fixed on a micrometer translation stage to adjust the length of the optical path. This beam is spatially filtered through a $35 \mu$m diameter pinhole and then expanded as to uniformly cover the CCD chip of the camera (512 $\times$ 512 pixels, Andor Luca R \textcopyright). The illumination beam is focalized on the sample with a plano-convex lens of focal length $12.5$ cm (waist diameter $\sim 200\, \mu \rm m$, laser intensity $\sim 250$ W/cm$^2$). The light scattered by the beads is collected in transmission with a dark-field reflecting objective (Edmund Optics \textcopyright, ReflX series) of ${\rm NA} \! = \! 0.5$ and $36 \rm X$ magnification. A small mask on the input of the objective limits the collection of light between ${\rm NA} \! = \! 0.2$ and ${\rm NA} \! = \! 0.5$, so the illumination beam is totally blocked after passing through the sample. This dark-field configuration prevents the saturation of the CCD chip. A non-polarizing $50/50$ beam splitter behind the objective combines the scattered light with the reference beam. The CCD camera records the interference pattern on $16\, {\rm bits}$ frames with a $22.5\, \rm Hz$ rate. The last beam-splitter is tilted by few degrees to be in off-axis configuration.

\section{Numerical reconstruction}
\subsection{Reconstruction of the scattering field}
\label{calcul}

In digital holography, the CCD sensor records an intensity $I_{ccd}$ which is the interference of the reference beam with the light scattered by the nano-objects. $I_{ccd}$ is thus a sum of 4 terms:

\begin{equation}
I_{ccd} = I_{ref} + I_{scatt} + E_{ref} \cdot E_{scatt}^* + E_{scatt} \cdot E_{ref}^*
\label{eq:intensity}
\end{equation}
where $I_{ref}$ is the intensity of the reference beam, $I_{scatt}$ the intensity of the scattered light and $E_{ref},E_{scatt}$ are the electric fields for the reference beam and the scattered light respectively. As the scattering cross section of a $100\, \rm nm$ diameter gold particle calculated with the Rayleigh-Mie scattering model is about $0.015\, {\rm \mu  m^2}$ at $660\, \rm nm$ \cite{vandeHulst}, the integration over the collection solid angle of the objective ($0.2 \leq NA \leq 0.5$) gives $I_{scatt} \simeq 3 \times 10^{-8} \times I_{illumination}$ for a single particle. Thus $I_{scatt}$ can be neglected compared to the other terms of Eq.(\ref{eq:intensity}). The field $E_{scatt}$, the amplitude of which is about $\sqrt{I_{scatt}}$, contains the phase of the scattered light necessary to a 3D localization of the particle. In the Fourier space, the tilted beam splitter adds a spatial frequency on the two conjugates terms of interference, and thus the different terms of $I_{ccd}$ are spatially separated, i.e. $I_{ref}$ remains centered on the zero frequency of the Fourier plane while the two cross terms of interference are centered on the spatial frequency induced by the off-axis geometry.

\begin{figure}[h]
\centering
\includegraphics[height=10 cm]{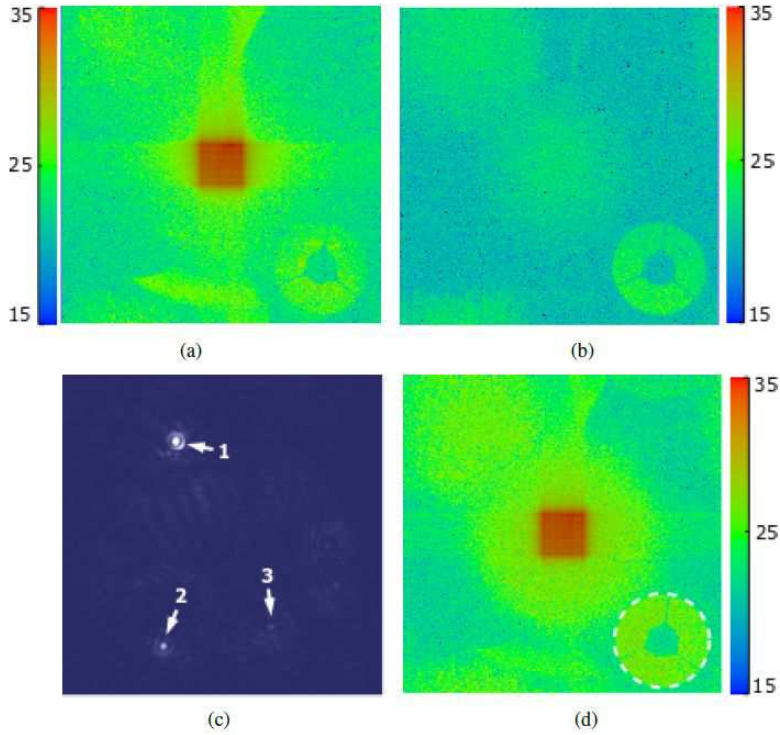}
\caption{a: Intensity $|\tilde{H}(k_x,k_y)|$ in logarithmic scale. The zero-order term appears as a red square distorted by the multiplication with the phase matrix $M$. The term of interest is located in the down-right corner. b: $|\tilde{H}(k_x,k_y)|$ when the average of the ten previous holograms is subtracted before calculating the FFT. The zero-order term is largely removed, so the recovery between this term and the region of interest in the down-right corner is reduced. c: Two-dimensional reconstruction at a fixed depth of the sample. A gold nanobead (1) is localized in this plan and the shape of the intensity of the field scattered by other beads at other depths (2 and 3) is visible. d: Intensity $|\tilde{H}(k_x,k_y)|$ in logarithmic scale when the sample is replaced by a diffusive paper. The area of the output pupil of the objective is sharply defined. The white-dotted circle shows the mask of the numerical filter used for the reconstruction.}
\label{fig:recons}
\end{figure}

Since 3D reconstruction is a time consuming task even using recent multi-cores processors, we developed parallel calculations on a Graphic card (Nvidia Geforce GT470, 448 cores) \cite{samson2011ol,ShimobabaSato2008,Ahrenberg2009,kang2009graphics}. Our \verb!C++! based algorithm uses the Nvidia CUDA library to decompose the calculations on the GPUs of the card. Among the existing methods to reconstruct holograms, the most common is the convolution method described by Schnars et al. \cite{schnars1994drh}. The main drawback of this method is that the pixel size of the reconstructed image depends on the distance of reconstruction, so that the image of a thick sample is distorted. Therefore this method is not convenient for 3D tracking, as the lateral scale depends on the depth of reconstruction considered. Here we chose to reconstruct the holograms with the angular spectrum method \cite{le2000numerical,yu2005wavelength}, which, by compensating the sphericity of the signal wave, allows to reconstruct the hologram without distortion. Our algorithm can be decomposed in 6 steps :
\begin{description}

 \item[i. \emph{subtraction of the background:}] in order to increase the signal-to-noise ratio, we subtract from the last recorded frame the average of the ten previous frames. Phase shifting holography \cite{yamaguchi1997psd} is also an effective technique for reducing noise, but the minimal delay $\Delta t=44$ ms between two frames, which is driven by our camera, is too long to use this technique for nanoparticle tracking. This step of calculation is performed only for particles in motion (i.e. we skipped this step for the results presented in \ref{fixed}).

 \item[ii. \emph{numerical correction of the signal wave sphericity:}] the hologram $I_{ccd}$ is multiplied by a complex phase matrix $M$ to compensate the sphericity induced by the microscope objective on the signal wave:
\begin{equation}
M(x,y,d)=\exp \left( \frac{i \pi (x^2 + y^2)}{\lambda d} \right),
\end{equation}
where $d$ is the local radius of curvature of the wave on the CCD plane. If the reference wave is a plane wave, this distance $d$ is also the distance between the CCD chip and the back focal plane of the objective.

 \item[iii. \emph{first FFT:}] the direct Fourier transform of the corrected hologram is calculated using the CUDA CUFFT library:
\begin{equation}
\tilde{H}(k_x,k_y)=FFT[ I_{ccd} \times M ].
\end{equation}
Fig. \ref{fig:recons}(a) shows the intensity $|\tilde{H}|$ in the \emph{k-space}, in logarithmic scale in the case where the background is not removed (step (i) skipped). In the middle of Fig. \ref{fig:recons}(a), the zero-order appears as a square because of the multiplication by the matrix $M$. The term related to $E_{scatt}$ is in the down-right corner,  centered on the spatial frequency induced by the off-axis geometry. The term related to $E_{scatt}^*$ is centered on the conjugate frequency (top-left corner). At this step, the calculation is equivalent of the reconstruction in one FFT (convolution method) of the hologram at the distance $d$ described in the previous step. Since the back focal plane of our microscope objective coincides with the output pupil plane, which is common for high magnification objective, we see on the down-right corner a sharp reconstruction of the output pupil plane. If we change in step (i) the parameter $d$ to $-d$, the image of the output pupil would be sharp in the top-left corner, while the term related to $E_{scatt}$ would be blurred.

Fig. \ref{fig:recons}(b) shows $|\tilde{H}|$ in the \emph{k-space} when the background is removed (step (i) performed). The zero-order term in the middle of the \emph{k-space} is largely removed compare to Fig. \ref{fig:recons}(a), which reduces the recovery between the zero-order and $E_{scatt}$.

\item[iv. \emph{spatial filtering and centering:}] to remove the zero-order term and replace the term related to $E_{scatt}$ in the middle of the Fourier plane, a round numerical filter which matches with the output pupil of the objective is applied. Since the shape of the pupil is sharp in the k-space, we can isolate precisely the pixels containing the signal, minimizing the loss of information. To more precisely calibrate the radius and the center of the filter, we used a diffusive paper as a sample before performing experiments. Since the paper scatters light uniformly, all the spatial frequencies that the microscope objective can collect are recorded. Fig. \ref{fig:recons}(d) shows the intensity $|\tilde{H}|$ in logarithmic scale when the sample is replaced by a diffusing paper. We clearly see the shape of the output pupil of the objective and the dark-field mask in the center. We set the filter mask to match with the shape of the output pupil (white dotted circle in Fig. \ref{fig:recons}(d)). The filtered part is then translated into the middle of a $512 \times 512$ calculation grid in order to compensate the off-axis shift.

\item[v. \emph{propagation: }]
the matrix obtained is multiplied by a propagation matrix $\tilde{K}(k_x,k_y,z)$, which is the exact form of the matrix propagation as given by Kim et al. \cite{kim2006interference}:
\begin{equation}
\tilde{K}(k_x,k_y,z)=\exp \left(i z \times \sqrt{k_0^2 - k_i^2 - k_j^2} \right)
\end{equation}
where
\begin{equation}
k_0=\frac{2 \pi n}{\lambda},\; \; k_x = \frac{2 \pi (x-256)}{512 \times \Delta_{pix}}, \; \; k_y = \frac{2 \pi (y-256)}{512 \times \Delta_{pix}}
\end{equation}
to propagate the hologram by a distance $z$ in the axial direction. $\Delta_{pix}$ is the magnified pixel size. These equations are suited for holograms of $512 \times 512$ pixels.

\item[vi. \emph{second FFT:}] finally the inverse FFT is calculated.
\end{description}

For each hologram, the steps (v) and (vi) are repeated in order to get a stack of the scattered field at different depths, with a propagation step  $\delta z = 100 \, \rm nm$:
\begin{equation}
H(x,y,n\! \cdot \! \delta z)=FFT^{-1}\left[ \tilde{H}(k_x,k_y) \times \tilde{K}(k_x,k_y, n\! \cdot \! \delta z) \right],
\end{equation}
where $n$ is an integer.

An example of reconstruction is shown in Fig. \ref{fig:recons}(c). We can see a nanoparticle (1) localized in the considered plane and the intensity of the field scattered by two particles at different depths (2 and 3). The reconstruction of a $50\, \mu \rm m$ thick volume requires to calculate one FFT to reconstruct the hologram in the output pupil plane, then to calculate 500 inverse FFTs for the slices of the stack. These 501 FFTs typically require one minute when the calculation is performed on the CPU, even with a recent multi-cores computer. We reduce this time by a factor of 30 when the calculation is parallelized on the graphic card.

\subsection{Method of localization}
\label{method}

\begin{figure}[h]
\centering
\includegraphics[width= 13 cm]{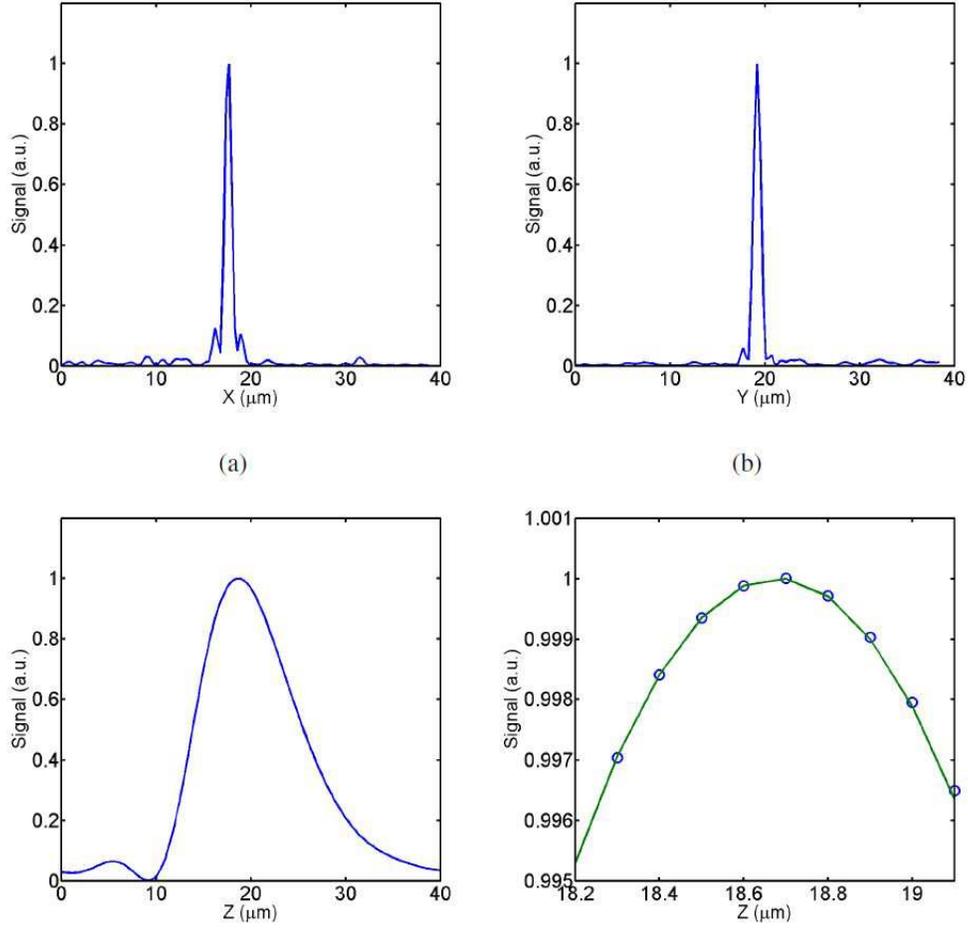}
%
\caption{PSF of the field's intensity for a $100\, \rm nm$ nanobead, (a) in $X$, (b) $Y$, (c) and in the $Z$ direction. (d) shows the top of the curve (c) (blue points) and the polynomial fit (green solid line).}
\label{fig:psf}
\end{figure}

Once the three dimensional map of the scattered field is calculated, the beads are localized by pointing the local maxima of the field's intensity. Fig. \ref{fig:psf} shows the field's intensity in the $X$,$Y$ and $Z$ directions. The Full Width at Half Maximum (FWHM) is about $1 \, \mu \rm m$ in lateral direction, and $14 \, \mu \rm m$ in the axial direction as expected for the Point Spread Function associated with our microscope objective\cite{nasse2010realistic,PSFlab}. In order to reach a sub-pixel size resolution, the intensity is fitted by a Gaussian curve in $X$ and $Y$ using the pixel for which the intensity is maximum and the two adjacent pixels. As the intensity profile in $Z$ is not gaussian and 10 times larger than in $X$ and $Y$, we fit the maximum of the peak with a parabola using the pixel $i$ for which the intensity is maximum and the two adjacent pixels $i-2$ and $i+2$ (Fig. \ref{fig:psf}(d)). We chose this simple localization method because programming a more elaborate fit, as T-Matrix theory based computation \cite{cheong2009flow}, with CUDA (programming language of the graphic card) is more complicated and would considerably slow down the process. Our method shows good performance (see Fig. \ref{fig:reso} and Fig. \ref{fig:error}), but a better resolution may be achieved using T-Matrix theory.

\section{Results}
\subsection{Embedded particles in a gel}\label{fixed}

\begin{figure}[h]
\addtocounter{subfigure}{0}
\subfigure[]{\includegraphics[width= 6 cm]{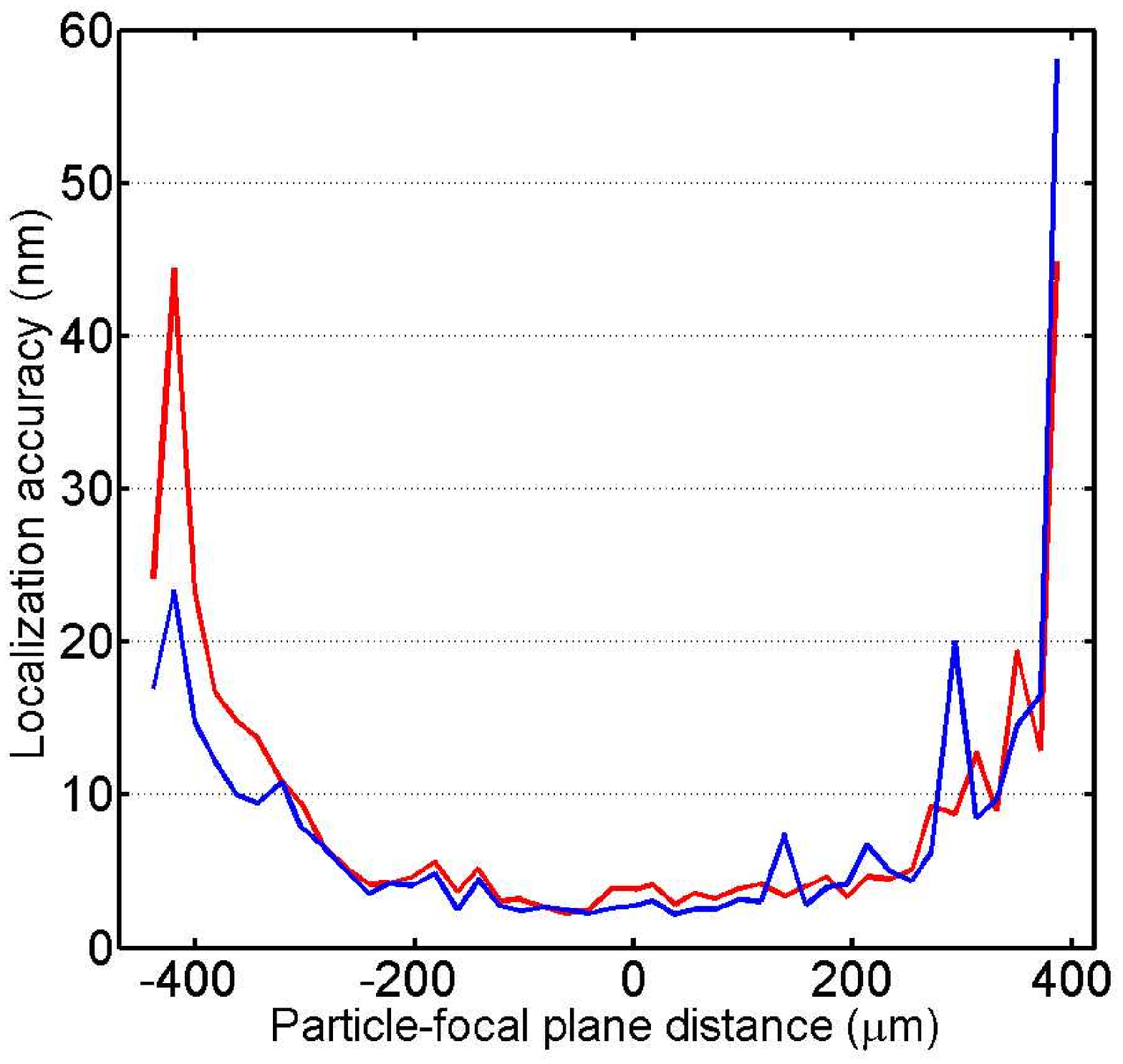}}
\hspace{1cm}
\subfigure[]{\includegraphics[width= 6 cm]{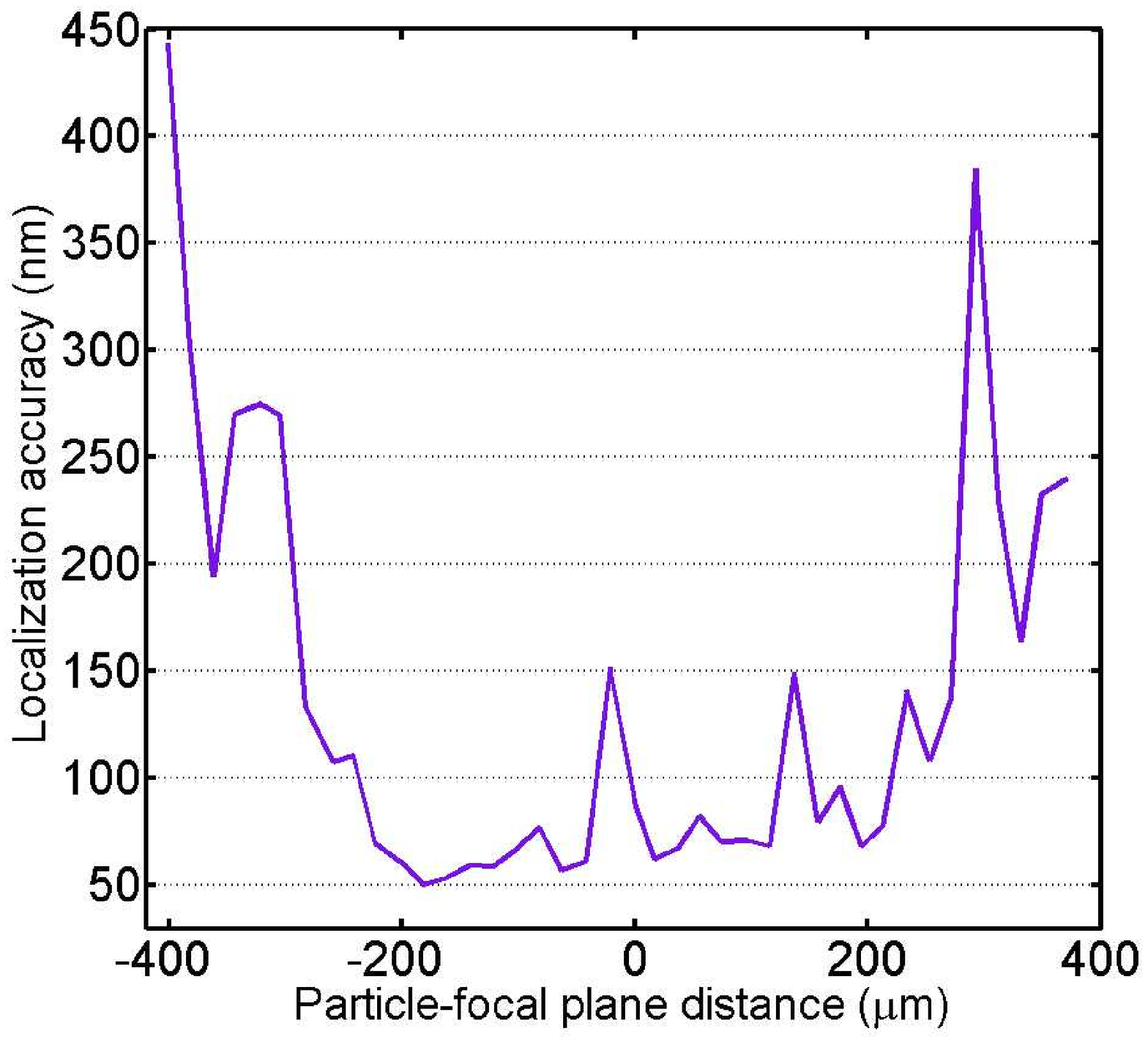}}
\caption{Localization accuracy as a function of the distance between the bead and the focal plane. (a) Lateral localization accuracy in $X$ (blue line) and $Y$ (red line). (b) Axial localization accuracy.}
\label{fig:reso}
\end{figure}

\begin{figure}
\centering
 \includegraphics[width= 9 cm]{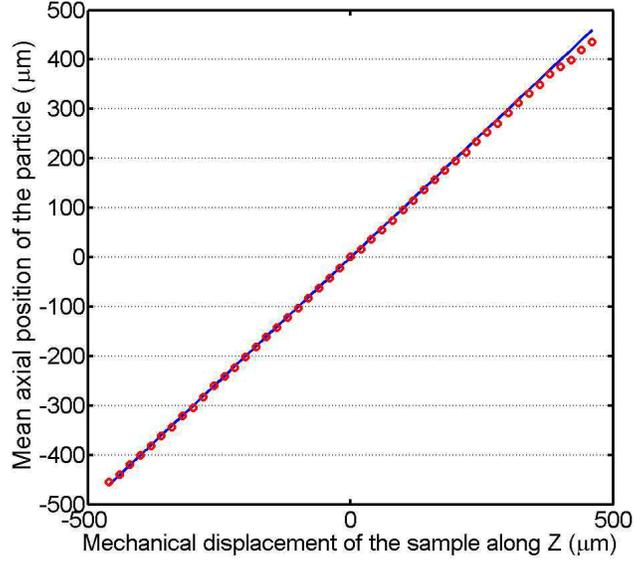}
 \caption{Mean axial position of an embedded particle, estimated using the reconstruction algorithm, as a function of the mechanical displacement of the sample along $Z$.}
 \label{fig:error}
\end{figure}

To evaluate the lateral and axial localization accuracy of our setup, we localize a single $100\, \rm nm$ diameter gold particle embedded in a 1\% agarose gel. The localization accuracy is evaluated by calculating the standard deviation of 200 positions of the bead obtained from successive frames, with an exposure time for each frame $\delta t = 1\, \rm ms$. For a particle in the focal plane of the objective, we found a lateral localization accuracy of $\sim 3\, \rm nm$ and an axial localization accuracy of $\sim 150\, \rm nm$ (Fig. \ref{fig:reso}(a) and \ref{fig:reso}(b)). Then the distance between the particle and the focal plane was increased by steps of $20 \, \mu \rm m$. For each step we recorded 200 holograms and determined the mean position of the bead as well as the lateral and axial accuracy. The Fig. \ref{fig:error} compares the mean axial position of the particle with the mechanical displacement along $Z$ of the sample, which are in excellent agreement. The localization accuracy in $X$,$Y$ and $Z$ as a function of the axial position of the bead is shown in Fig. \ref{fig:reso}(a) and \ref{fig:reso}(b). While the lateral accuracy is constant ($\sim 3\, \rm nm$) for $|Z| < 250$ nm, the axial accuracy slightly depends on $Z$. It is about $\sim 150\, \rm nm$ around $Z=0$, then decreases to $\sim 70\, \rm nm$ for $|Z| < 250$ nm. This accuracy strongly increases for $|Z| > 250\, \mu \rm m$, and for $Z>400\, \mu \rm m$, the localization of the particle is not possible because the scattered signal level reaches the noise level. The local maximum at $Z=0$ observed on the axial accuracy curve (Fig. \ref{fig:reso}(b)) shows that the localization is not optimal when the gold particle is the focal plane of the objective. In this case the particle is imaged on a small area of the CCD chip, so that the interference pattern spreads on a small number of pixels, which degrades the quality of the reconstruction \cite{fournier2010single}.

In the case of particle tracking, the position of the focal plane in the sample has to be fixed. To minimize the spherical aberrations due to the presence of the coverslip, the focal plane should coincide with the sample-coverslip interface. In this configuration, a moving particle cannot cross the focal plane, so that the localization accuracy remains optimal and constant over a depth of $250\, \mu \rm m$, that corresponds to the part between abscissa 0 and abscissa 250 in Fig. \ref{fig:reso}(a) and \ref{fig:reso}(b). Yet tracking particles remains possible when the focal plane is set above the sample-coverslip interface as demonstrated below, as the cost of a slightly worse axial localization accuracy.

\subsection{Particles in Brownian motion}

We now consider gold particles in Brownian motion. According to the Stokes-Einstein equation, the diffusion coefficient $D$ of the particles is given by:
\begin{equation}
D=\frac{k_{\mathrm{B}}T}{6\pi\eta r} = 4.2 \pm 0.2 \, \mu \rm m^2 \,\rm s^{-1},
\label{Stokes}
\end{equation}
where $k_{\mathrm{B}}$ is the Boltzmann constant, $T$ the room temperature ($20\,^{\circ}\rm C$), $\eta$ the viscosity of water (1.0\,{\rm mPa.s} at $20\,^{\circ}\rm C$) and $r = 50 \pm 2$ nm the radius of the nanobead (size dispersity given by the provider BBInternational). The exposure time is $\delta t = 1\, \rm ms$ and the time between two frames $\Delta t = 44\, \rm ms$. The mean distance traveled along one direction by a Brownian particle during $\delta t$ is $90\, \rm nm$, which is smaller than the lateral size of the magnified pixels ($160\, \rm nm$). Consequently, the signal from a particle is not blurred over several pixels during $\delta t$. The mean distance covered along a given axis during $\Delta t$ is $620\, \rm nm$, which corresponds to approximately $5$ pixels.

\begin{figure}[h]
\centering
\includegraphics[height = 7.5 cm, width = 8.0 cm]{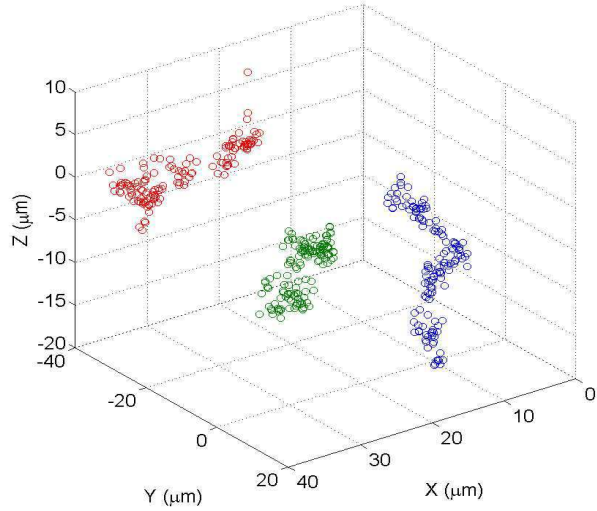}
\caption{3D trajectories of 3 particles (red, green, blue) reconstructed from 200 successive frames. The focal plane ($Z=0$) was set at about $150\, \mu$m above the coverslip. Although this setting is not optimal for tracking a particle diffusing around the focal plane, as explained in section \ref{fixed}, trajectory reconstruction is still possible (red trajectory).}
\label{fig:motion}
\end{figure}

\begin{figure}[h]
\centering
\includegraphics[width = 12.0 cm]{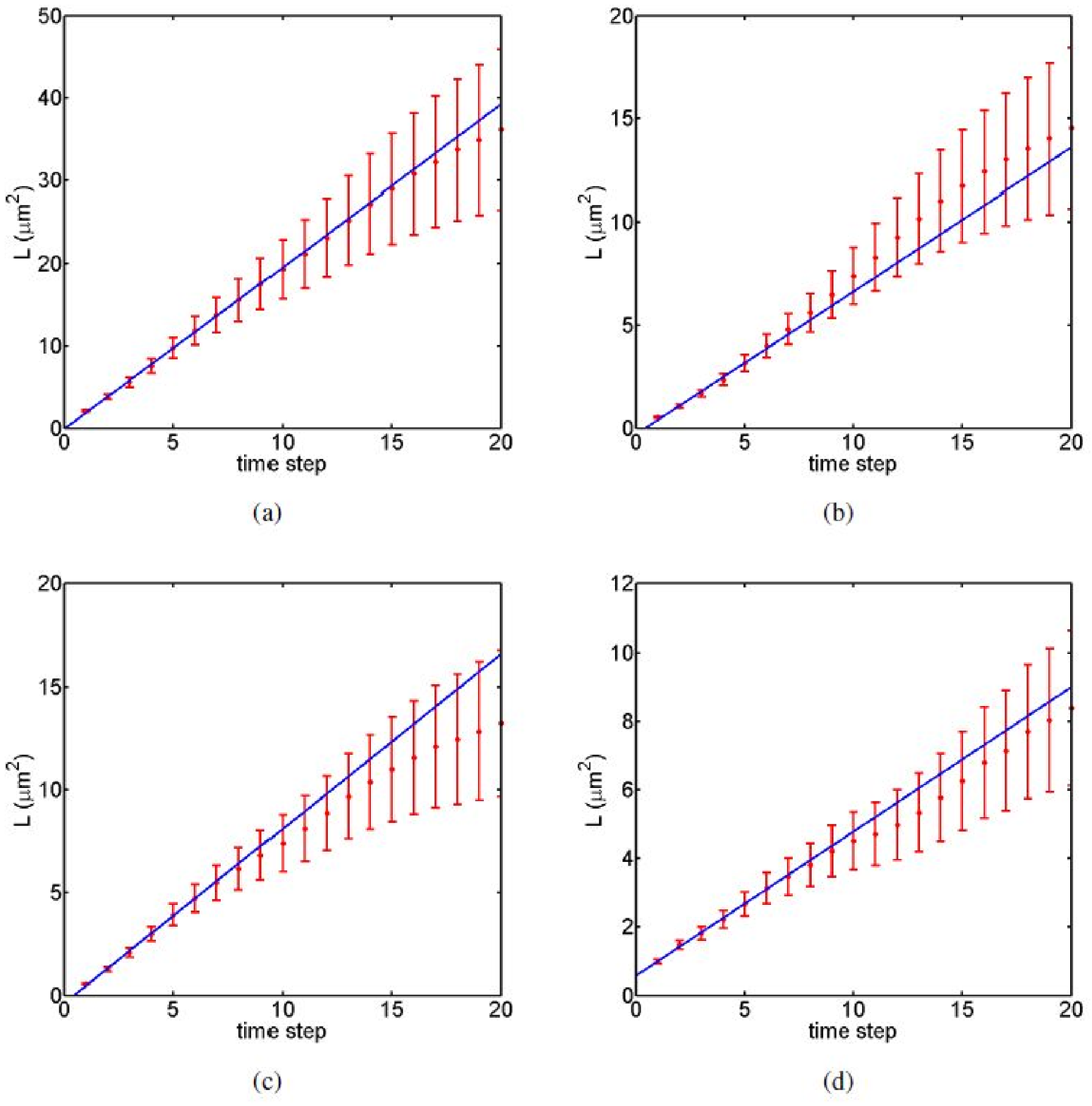}
\caption{Mean Square Displacement curves derived from the trajectory of a nanoparticle in Brownian motion (200 points, time step $\Delta t$ = 44 ms). (a) 3D MSD, (b) MSD along $X$, (c) MSD along $Y$ and (d) MSD along $Z$. Blue line: linear fit over the first 6 points of the MSD.}
\label{fig:msd}
\end{figure}
As shown in Fig. \ref{fig:motion}, our method allows us to simultaneously track several particles. A volume of $80 \times 80 \times 250\, \mu \rm m$ ($X \times Y \times Z$), i.e. $512 \times 512\times 400$ pixels, can be reconstructed from a single hologram, and the localization method described in the section \emph{\ref{method}} can be performed to localize several beads with a sub-pixel accuracy. By repeating the algorithm for successive frames, we could for instance reconstruct the trajectories of 3 gold particles diffusing in water (Fig. \ref{fig:motion}). We were able to track particles during up to 10 s ($\sim 200$ frames). Since the time needed to reconstruct a volume of $512 \times 512\times 400$ pixels is about 0.5 s (i.e. much larger than $\Delta t = 44$ ms), reconstruction is necessarily a post-processing process in this case.

In contrast, when a single particle is tracked, 3D localization in a given frame requires to reconstruct only a few slices around the position of the particle in the previous frame. For a $100\, \rm nm$ particle, the mean distance traveled along $Z$ during $\Delta t$ is $\sim 600 \, \rm nm$, thus only 24 reconstructions (12 reconstructions above/below the previous position of the particle) are sufficient to find the new position of the particle. The calculation of these 24 slices from the hologram requires $33 \, \rm ms$, which is smaller than $\Delta t = 44$ ms. Real-time tracking is thus possible for a single particle under the condition that reconstruction is performed fast enough, for instance by using a graphic card unit as described above.

To evidence the Brownian motion, we calculated the 3D Mean Square Displacement (MSD) of a bead from the red trajectory plotted in Fig. \ref{fig:motion}:
\begin{equation}
{\rm MSD}(n\cdot \Delta t)=\frac{1}{N-n} \left( \sum_{i=1}^{n-N} (x_{i+n} -
x_i)^2 + (y_{i+n} - y_i)^2 + (z_{i+n} - z_i)^2 \right),
\end{equation}
where $N = 200$ is the total number of positions, as well as the 1D MSD along the directions $X$, $Y$, and $Z$. For a Brownian motion with a diffusion constant $D$, the MSD curve depends linearly on time, and the slope of the curve is $2^{n_d} \, D$, where $n_d \! =\! 3$ for the 3D MSD (Fig. \ref{fig:msd}(a)) and $n_d \! =\! 1$ for 1D MSD (Fig. \ref{fig:msd}(b), (c) and (d)). As expected for a Brownian motion, experimental 3D and 1D MSD depend linearly on time, and a fit of the first six points of the curves gives $D = 4.3 \pm 0.5\, \mu \rm m^2\rm s^{-1}$ along $X$, $D = 5.0 \pm 0.3\, \mu \rm m^2\rm s^{-1}$ along $Y$, $D = 2.9 \pm 0.5\, \mu \rm m^2\rm s^{-1}$ along Z and $D = 4.1 \pm 0.5\, \mu \rm m^2\rm s^{-1}$ for the 3D motion. These values are in agreement with the theoretical value $D = 4.2\, \pm 0.2 \, \mu \rm m^2\rm s^{-1}$ predicted by Eq. (\ref{Stokes}).

\section{Conclusion}
In this paper, we show that Digital Holographic Microscopy can be used to track $d=100\, \rm nm$ gold particles diffusing in water. As the intensity of the light scattered by such nanoparticles is height orders of magnitude smaller than that of the excitation light, we combined holography to standing wave free dark-field microscopy to completely block the illumination beam, thereby preventing a saturation of the CCD chip of the camera. A single hologram, recorded with an exposure time of only 1 ms, is sufficient to localize several particles in a 250 $\mu$m thick sample, with a lateral ($x,y$) localization accuracy of $\sim 3\, \rm nm$ and an axial ($z$) localization accuracy of $\sim 70\, \rm nm$. As our dark field microscope involves a NA=0.5 reflecting microscope objective, the resolution, especially in $z$, obtained here with nanoparticles ($d=100$ nm), is lower than the resolution reached in in-line holography using a NA=1.4 objective and micron-sized objects ($d=1.5~\mu$m) \cite{cheong2009holographic}. This is the cost to pay for the detection of nanometer-sized particles.
We were able to reconstruct particle trajectories, evidence the Brownian nature of the motion and determine the related diffusion coefficient. The accuracy achieved by our setup is comparable with that reached with super-resolution microscopy: a localization accuracy of $20$ nm in X and Y, and $50$ nm in Z has been reported for fluorescence dye or quantum dots imaged with STORM microscopy using optical astigmatism \cite{huang2008three}, and an accuracy of $10$ nm over a depth of $2\, \mu$m has been reached using a double helix PSF \cite{pavani2009three} . Severals results of 3D tracking of quantum dots using PALM microscopy report an axial accuracy between $10$ and $75$ nm, over a depth of typically $1\, \mu$m \cite{juette2008three,ram2008high}. PALM and STORM microscopy reach higher 3D localization accuracies than our DHM setup, but over a depth two order of magnitude lower. This ability to track nanoparticles up to $250\, \mu$m from the focal plane with a constant localization accuracy is the strength of DHM compare to super-resolution techniques. Our results pave the way for the use of gold nanobeads as markers in more complex media such as cellular environment.

\section*{Acknowledgments}
This work was supported by funds from the French National Research Agency (ANR SIMI 10 and ANR 3D BROM), Centre National de la Recherche Scientifique (CNRS) and \'Ecole normale sup\'erieure (ENS).
The authors thank Mathieu Coppey, Fred Etoc and Jasmina Dikic for their suggestions and careful reading.

\begin{thebibliography}{99}

\bibitem{huang2008three}
B. Huang, W. Wang, M. Bates, and X. Zhuang,
``Three-dimensional super-resolution imaging by stochastic optical reconstruction microscopy,'' Science {\bf 319,} 810 (2008).

\bibitem{bewersdorf2006h2ax}
J. Bewersdorf, B.T. Bennett, and K.L. Knight,
``H2AX chromatin structures and their response to DNA damage revealed by 4Pi microscopy,'' Proc. Nat. Acad. Sci. USA {\bf 103,} 18137 (2006).

\bibitem{pavani2009three}
S.R. Pavani, M.A. Thompson, J.S. Biteen, S.J. Lord, N. Liu, R.J. Twieg, R. Piestun, and W.E. Moerner,
``Three-dimensional, single-molecule fluorescence imaging beyond the diffraction limit by using a double-helix point spread function,'' Proc. Nat. Acad. Sci. USA {\bf 106,} 2995 (2009).

\bibitem{juette2008three}
M.F Juette, T.J. Gould, M.D. Lessard, M.J. Mlodzianoski, B.S. Nagpure, B.T. Bennett, S.T. Hess, and J. Bewersdorf,
``Three-dimensional sub--100 nm resolution fluorescence microscopy of thick samples,'' Nature Methods {\bf 5,} 527--529 (2008).

\bibitem{ram2008high}
S. Ram, P. Prabhat, J. Chao, E.S. Ward, and R.J. Ober,
``High Accuracy 3D Quantum Dot Tracking with Multifocal Plane Microscopy for the Study of Fast Intracellular Dynamics in Live Cells,'' Biophysical Journal {\bf 95,} 6025 (2008).

\bibitem{schnars1994drh}
U. Schnars and W. J{\"u}ptner,
``Direct recording of holograms by a CCD target and numerical reconstruction,'' \ao {\bf 33,} 179--181 (1994).

\bibitem{leith1965}
E. Leith and J. Upatniek,
``Wavefront Reconstruction Photography,'' Physics Today {\bf 18,} 26 (1965).

\bibitem{vandeHulst}
H.C. van de Hulst,
``Light Scattering by Small Particles,'' (Dovers Publications Inc, 1957).

\bibitem{cheong2009flow}
F.C. Cheong, B. Sun, R. Dreyfus, J. Amato-Grill, K. Xiao, L. Dixon, and D.G. Grier,
``Flow visualization and flow cytometry with holographic video microscopy,'' \opex {\bf 17,} 13071--13079 (2009).

\bibitem{cheong2009holographic}
F.C. Cheong, S. Duarte, S.H. Lee, and D.G. Grier,
``Holographic microrheology of polysaccharides from Streptococcus mutans biofilms'' Rheologica Acta {\bf 48,} 109--115 (2009).

\bibitem{cheong2010strategies}
F.C. Cheong, B.J. Krishnatreya, and D.G. Grier,
``Strategies for three-dimensional particle tracking with holographic video microscopy,'' \opex {\bf 18,} 13563--13573 (2010).

\bibitem{xu2003tracking}
W. Xu, M.H. Jericho, H.J. Kreuzer, and I.A. Meinertzhagen,
``Tracking particles in four dimensions with in-line holographic microscopy,'' \ol {\bf 28,} 164--166 (2003).

\bibitem{sheng2006digital}
J. Sheng, E. Malkiel, and J. Katz,
``Digital holographic microscope for measuring three-dimensional particle distributions and motions,'' \ao {\bf 45,} 3893--3901 (2006).

\bibitem{lee2007characterizing}
S.H Lee, Y. Roichman, G.R. Yi, S.H. Kim, S.M. Yang, A. Van Blaaderen, P. Van Oostrum, and D.G. Grier,
``Characterizing and tracking single colloidal particles with video holographic microscopy,'' \opex {\bf 15,} 18275--18282 (2007).

\bibitem{speidel2009interferometric}
M. Speidel, L. Friedrich, and A.Rohrbach,
``Interferometric 3D tracking of several particles in a scanning laser focus,'' \opex {\bf 17,} 1003--1015 (2009).

\bibitem{leithjosa}
E. Leith and J. Upatnieks,
``Microscopy by wavefront reconstruction,'' \josa {\bf 55,} 569-–570 (1965).

\bibitem{cuche2000spatial}
E. Cuche, P. Marquet, and C. Depeursinge,
``Spatial filtering for zero-order and twin-image elimination in digital off-axis holography,'' \ao {\bf 39,} 4070--4075 (2000).

\bibitem{gross2007digital}
M. Gross and M. Atlan,
''Digital holography with ultimate sensitivity'' \ol {\bf 32,} 909--911 (2007)

\bibitem{jain2006calculated}
P.K. Jain, K.S. Lee, I.H. El-Sayed, and M.A. El-Sayed,
``Calculated absorption and scattering properties of gold nanoparticles of different size, shape, and composition: applications in biological imaging and biomedicine,''
J. Phys. Chem. B {\bf 110,} 7238-–7248 (2006).

\bibitem{atlan2008hhm}
M. Atlan, M. Gross, P. Desbiolles, {\'E}. Absil, G. Tessier, and M. Coppey-Moisan,
``Heterodyne holographic microscopy of gold particles,'' \opex {\bf 33,} 500--502 (2008).

\bibitem{cognet2003single}
L. Cognet, C. Tardin, D. Boyer, D. Choquet, P. Tamarat, and B. Lounis,
``Single metallic nanoparticle imaging for protein detection in cells,'' Proc. Nat. Acad. Sci. USA {\bf 100,} 11350 (2003).

\bibitem{warnasooriya2010imaging}
N. Warnasooriya, F. Joud, F. Bun, G. Tessier, M. Coppey-Moisan, P. Desbiolles, M. Atlan, M. Abboud, and M. Gross,
``Imaging gold nanoparticles in living cell environments using heterodyne digital holographic microscopy,'' \opex {\bf 18,} 3264--3273 (2010).

\bibitem{shaffer2010real}
E. Shaffer, P. Marquet, and C. Depeursinge,
``Real time, nanometric 3D-tracking of nanoparticles made possible by second harmonic generation digital holographic microscopy,'' \opex {\bf 18,} 17392--17403 (2010).

\bibitem{verpillat2010digital}
F. Verpillat, F.  Joud, M. Atlan, and M. Gross,
''Digital holography at shot noise level'' Journal of Display Technology {\bf 6,} 455--464 (2010)

\bibitem{dubois2008dark}
F. Dubois and P. Grosfils,
''Dark-field digital holographic microscopy to investigate objects that are nanosized or smaller than the optical resolution'' \ol {\bf 33,} 2605--2607 (2008)

\bibitem{samson2011ol}
B. Samson, F. Verpillat, M. Gross, and M. Atlan,
``Video-rate laser Doppler vibrometry by heterodyne holography,'' \ol {\bf 36,} 1449--1451 (2011).

\bibitem{ShimobabaSato2008}
T. Shimobaba, Y. Sato, J. Miura, M. Takenouchi, and T. Ito,
``Real-time digital holographic microscopy using the graphic processing unit,'' \opex {\bf 16,} 11776--11780 (2008).

\bibitem{Ahrenberg2009}
L. Ahrenberg, A.J. Page, B.M. Hennelly, J.B. McDonald, and T.J. Naughton,
``Using commodity graphics hardware for realtime digital hologram view-reconstruction,'' J. Display Technol. {\bf 5,} 111 (2009).

\bibitem{kang2009graphics}
H. Kang, F. Yara{\c{s}}, and L. Onural,
``Graphics processing unit accelerated computation of digital holograms,'' \ao {\bf 38,} 137--143 (2009).

\bibitem{le2000numerical}
F. Le Clerc, L. Collot, and M. Gross,
``Numerical heterodyne holography with two-dimensional photodetector arrays,'' \ol {\bf 25,} 716--718 (2000).

\bibitem{yu2005wavelength}
L. Yu and M.K. Kim,
``Wavelength-scanning digital interference holography for tomographic three-dimensional imaging by use of the angular spectrum method,'' \ol {\bf 30} 2092--2094 (2005).

\bibitem{yamaguchi1997psd}
I. Yamaguchi and T. Zhang,
``Phase-shifting digital holography,'' \opex {\bf 22,} 126--1270 (1997).

\bibitem{kim2006interference}
M.K. Kim, L. Yu, and C.J. Mann,
``Interference techniques in digital holography,'' Journal of Optics A: Pure and Applied Optics {\bf 8,} S518--S523 (2006).

\bibitem{nasse2010realistic}
M. J. Nasse and J. C. Woehl,
``Realistic modeling of the illumination point spread function in confocal scanning optical microscopy,'' \josa A {\bf 27,} 295--302 (2010).

\bibitem{PSFlab}
PSF Lab, http://onemolecule.chem.uwm.edu/software.

\bibitem{fournier2010single}
C. Fournier, L. Denis, and T. Fournel,
``On the single point resolution of on-axis digital holography,'' \josa A {\bf 27,} 1856--1862 (2010).

\end{thebibliography}
\end{document}